\documentclass[reprint, 10pt, a4paper, aip, floatfix]{revtex4-1}
\usepackage[utf8]{inputenc}
\usepackage{amsmath}
\usepackage{amsfonts}
\usepackage{amssymb}
\usepackage{mathrsfs}
\usepackage{subfigure}
\usepackage[hidelinks]{hyperref}
\usepackage{outlines}
\usepackage{tikz}

\newcommand{\bv}{\mathbf{v}}
\newcommand{\bE}{\mathbf{E}}
\newcommand{\bB}{\mathbf{B}}
\newcommand{\bF}{\mathbf{F}}
\newcommand{\bR}{\mathbf{R}}

\newcommand{\tr}{\tilde{r}}
\newcommand{\tn}{\tilde{n}}
\newcommand{\tT}{\tilde{T}}
\newcommand{\tB}{\tilde{B}}

\usepackage{tensor}
\newcommand{\te}{\tensor}
\newcommand{\beq}{\begin{equation}}
\newcommand{\eeq}{\end{equation}}

\newcommand{\bvec}{\begin{pmatrix}}
\newcommand{\evec}{\end{pmatrix}}
\newcommand{\lp}{\left(}
\newcommand{\rp}{\right)}

\newcommand{\pa}[2]{\frac{\partial #1}{\partial #2}}

\newcommand{\teps}{\tilde{\epsilon}}

\begin{document}

\title{Radial Current and Rotation Profile Tailoring in Highly Ionized Linear Plasma Devices}
\date{\today}
\author{E. J. Kolmes}
\thanks{Co-first author}
\email[Electronic mail: ]{ekolmes@princeton.edu}
\author{I.~E.~Ochs}
\thanks{Co-first author}
\email[Electronic mail: ]{iochs@princeton.edu}
\affiliation{Department of Astrophysical Sciences, Princeton University, Princeton, New Jersey, USA}
\author{M.~E.~Mlodik}
\affiliation{Department of Astrophysical Sciences, Princeton University, Princeton, New Jersey, USA}
\author{J.-M.~Rax}
\affiliation{Universit\'{e} de Paris XI - \'{E}cole Polytechnique, LOA-ENSTA-CNRS, 91128 Palaiseau, France}
\author{R.~Gueroult}
\affiliation{LAPLACE, Universit\'{e} de Toulouse, CNRS, INPT, UPS, 31062 Toulouse, France}
\author{N.~J.~Fisch}
\affiliation{Department of Astrophysical Sciences, Princeton University, Princeton, New Jersey, USA}

\begin{abstract}
In a rotating magnetized plasma cylinder with shear, cross-field current can arise from inertial mechanisms and from the cross-field viscosity. Considering these mechanisms, it is possible to calculate the irreducible radial current draw in a cylindrical geometry as a function of the rotation frequency. The resulting expressions raise novel possibilities for tailoring the electric field profile by controlling the density and temperature profiles of a plasma. 
\end{abstract}

\maketitle

\section{Introduction}

In many technologies that rely on rotating plasmas,\cite{Lehnert1971} the cross-field conductivity is of great practical concern. 
Usually, the rotation is produced by imposing an electrical potential difference perpendicular to a magnetic field; the resulting fields and power dissipation depend critically on the plasma conductivity.
However, a large cross-field conductivity drives up the cost of maintaining $\bE \times \bB$ rotation. 
For instance, the viability of mass filters based on rotating plasmas as a replacement for chemical separation techniques is partly dependent on power efficiency.\cite{hellsten1977balance,ohkawa2002band,fetterman2011magnetic,Gueroult2015, Gueroult2018ii, Zweben2018}
Other examples include fusion schemes that rely on rotating plasmas,\cite{Bekhtenev,Ellis2001,fetterman2008alpha,Kolmes2018Strategies,Rax2017} where the efficiency is similarly limited by the cross-field conductivity. 

In some of these applications, it is required to have very specific control over the rotation profile (i.e.~the shear). 
Achieving this control is a nontrivial problem; past efforts in linear devices have often focused on biasable annular end electrodes.\cite{Tsushima1986, Shinohara2001, Zhou2012, Schaffner2012, Gueroult2016, desjardins2016dynamics, Jin2019, Poulos2019}
Another possibility is through waves, or stationary perturbations,\cite{fetterman_rot} though this has not yet been attempted experimentally.

A recent paper by Rax \textit{et al.}\cite{Rax2019} analyzed the cross-field conductivity in a plasma undergoing solid-body rotation. 
The present paper addresses general rotation profiles, supplementing Rax \textit{et al.} \cite{Rax2019}  as well as previous considerations of plasma 	conductivity in a rotating plasma with general rotation profiles, where the effects of shear have also been addressed.\cite{HelanderSigmar,Rozhansky2008}
In particular, we find that sheared rotation modifies the form of the inertial conductivity discussed by Rax \textit{et al.} and results in additional viscosity-dependent effects.
Importantly, the radial current calculation presented here suggests that control over the radial temperature profile could help to control the rotation profile even without the use of end electrodes. 

The paper is organized as follows:  In Sec.~\ref{sec:derivation}, we derive the basic drift equations for ions and electrons in a rotating plasma.  We then proceed to order terms based on smallness parameters.  
The two smallness parameters rely on the ion-electron collision frequency and the plasma rotation frequency being small compared to the gyrofrequency.
We also assume that radial pressure gradients are small compared to the radial electric forces. 
In Sec.~\ref{sec:nondimensionalization}, we write the governing equations of the system in nondimensional form and identify the dimensionless parameters that determine their behavior. 
In Sec.~\ref{sec:mechanisms}, we identify the contributions to the radial current arising from plasma viscosity, inertia, and particle sources, respectively.
We apply these relations in  Sec.~\ref{sec:profile} to show how sources and sinks might be used to control the rotation profile.
In Sec.~\ref{sec:discussion}, we compare the results derived  here to previous derivations in the literature, and we discuss how alternate viscosity models, for example those used to describe rotating turbulent plasma,  might be incorporated. 
In Appendix A, we provide two derivations of the Braginskii viscosity components in a rotating plasma.

\section{Classical Cross-Field Conductivity in a Rotating Plasma with Shear} \label{sec:derivation}

Consider the two-fluid momentum equations for a plasma with a single ion species of charge $Z e$ and mass $m_i$:
\begin{align}
&\frac{\partial}{\partial t} \big( m_i n_i \bv_i \big) + \nabla \cdot (m_i n_i \bv_i \bv_i) = Z e n_i (\bE + \bv_i \times \bB) \nonumber \\
&\hspace{70 pt}- \nabla p_i - \nabla \cdot \pi_i + \bR_i + m_i s_i \, \bv_{i}^\text{src} \label{eqn:ionMomentum} \\
&\frac{\partial}{\partial t} \big( m_e n_e \bv_e \big) + \nabla \cdot (m_e n_e \bv_e \bv_e) = - e n_e (\bE + \bv_e \times \bB) \nonumber \\
&\hspace{70 pt}- \nabla p_e - \nabla \cdot \pi_e + \bR_e + m_e s_e \, \bv_{e}^\text{src}. \label{eqn:electronMomentum}
\end{align}
Here $m_e$ is the electron mass, $p_{i \backslash e} = n_{i \backslash e} T_{i \backslash e}$ are the scalar pressures, $\pi_{i \backslash e}$ are the viscosity tensors, and $\bR_i$ and $\bR_e$ are the ion-electron and electron-ion friction force densities. If there are sources or sinks of sinks of charged particles (corresponding e.g. to neutral ionization), their source rates are denoted by $s_{i \backslash e}$ and the velocities of the particles being added or removed are denoted by $\bv_{i \backslash e}^\text{src}$. Using the density continuity equation, the LHS of Eqs.~(\ref{eqn:ionMomentum}) and (\ref{eqn:electronMomentum}) can be rewritten as 
\begin{align}
&\frac{\partial}{\partial t} \big( m_s n_s \bv_s \big) + \nabla \cdot (m_s n_s \bv_s \bv_s) \rightarrow m_s n_s \frac{d\bv_s}{dt} + m_s \bv_s s_s, 
\end{align}
where $d/dt$ is the advective derivative $\partial/\partial t + \bv_s \cdot \nabla$. 

Eqs.~(\ref{eqn:ionMomentum}) and (\ref{eqn:electronMomentum}) can be rearranged, after taking the cross product with $\hat b = \bB / B$ on both sides, to get expressions for the components of $\bv_{i \backslash e}$ that are perpendicular to $\hat b$:
\begin{align}
\bv_{i \perp} &= \frac{\bE \times \bB}{B^2} - \frac{(\nabla p_i + \nabla \cdot \pi_i) \times \hat b}{n_i m_i \Omega_i} + \frac{\bR_i \times \hat b}{n_i m_i \Omega_i} \nonumber \\
&\hspace{10 pt}- \frac{1}{\Omega_i} \frac{d \bv_i}{dt} \times \hat b + \frac{s_i (\bv_{i}^\text{src} - \bv_i) \times \hat b}{n_i \Omega_i}
\end{align}
\begin{align}
\bv_{e \perp} &= \frac{\bE \times \bB}{B^2} - \frac{(\nabla p_e + \nabla \cdot \pi_e) \times \hat b}{n_e m_e \Omega_e} + \frac{\bR_e \times \hat b}{n_e m_e \Omega_e} \nonumber \\
&\hspace{10 pt}- \frac{1}{\Omega_e} \frac{d \bv_e}{dt} \times \hat b + \frac{s_e (\bv_{e}^\text{src} - \bv_e) \times \hat b}{n_e \Omega_e}, 
\end{align}
where $\Omega_i \doteq Z_i e B / m_i$ and $\Omega_e \doteq - e B / m_e$. 
Starting from the left, these flows are the $\bE \times \bB$ drift, the diamagnetic drift, the viscosity drift, the frictional drift, the polarization drift, and a drift due to particle injection whose physical origin is analyzed in Sec.~\ref{subsec:injectionCurrent}. 
If $|\Omega_e| \gg \nu_{ei}$, the friction force densities are:\cite{Braginskii1965}
\begin{align}
\bR_i &= n_i m_i \nu_{ie} (\bv_e - \bv_i) - \frac{3 n_e \nu_{ei} \nabla T_e \times \hat b}{2 \Omega_e} \\
\bR_e &= n_e m_e \nu_{ei} (\bv_i - \bv_e) + \frac{3 n_e \nu_{ei} \nabla T_e \times \hat b}{2 \Omega_e}, 
\end{align}
where $\nu_{ie}$ and $\nu_{ei}$ are the ion-electron and electron-ion collision frequencies. 

Consider an axisymmetric plasma with $\bB = B \hat z$, $\bE = E \hat r$, and all gradients in the $\hat r$ direction. In steady state, the velocities in the $\hat r$ and $\hat \theta$ directions can be written as 
\begin{gather}
v_{sr} = \frac{R_{s \theta} - (\nabla \cdot \pi_s)_\theta + m_s s_s (v_{s \theta}^\text{src} - v_{s \theta})}{n_s m_s \Omega_s [1 + (r v_{s \theta})'/ r \Omega_s]} \label{eqn:vr}
\end{gather}
and 
\begin{align}
v_{s \theta} &= - \frac{E}{B} + \frac{p_s'}{n_s m_s \Omega_s} + \frac{(\nabla \cdot \pi)_r}{n_s m_s \Omega_s} - \frac{R_{sr}}{n_s m_s \Omega_s} \nonumber \\
&\hspace{40 pt}- \frac{s_s (v_{sr}^\text{src} - v_{sr})}{n_s \Omega_s} - \frac{v_{sr} v_{sr}'}{\Omega_s} + \frac{v_{s \theta}^2}{r \Omega_s} \, , \label{eqn:vtheta}
\end{align}
where the prime denotes a derivative with respect to $r$.
The bracketed part of the denominator of Eq.~(\ref{eqn:vr}) and the last two terms of Eq.~(\ref{eqn:vtheta}) come from the advective part of the total derivative $d \bv_s / dt$, and correspond to the fictitious forces in a rotating frame.

In a strongly magnetized plasma, the ion-electron collision frequency is much smaller than the ion gyrofrequency, so $\epsilon \doteq \nu_{ie} / \Omega_i$ is a small parameter. In many rotating plasmas of interest, the ratio $\delta \doteq E / (r B \Omega_i)$ of the rotation frequency to the gyrofrequency is also small.
Consider a double ordering in $\delta$ and $\epsilon$, similar to the one used by Spitzer,\cite{Spitzer1952} where we denote a term of order $\delta^n \epsilon^m$ by the tuple $(n,m)$, as in $v_{i \theta}^{(n,m)}$. 

Assume that $p' / n e E \sim \mathcal{O}(\delta)$. This assumption is convenient, and it is reasonable in many cases,\cite{Lehnert1971, Rax2017} but an analogous calculation can be carried out for a stronger or weaker pressure force. It will also be useful to pick an ordering for the viscous forces. This will be discussed in greater detail later on, but for now, take the $\hat r$ component of $(\nabla \cdot \pi_s) / e n_s E$ to be $\mathcal{O}(\delta^2)$ and the $\hat \theta$ component to be $\mathcal{O}(\delta^2 \epsilon)$. Finally, since the continuity equation suggests $v_{sr} \sim R s_s / n_s$ for a characteristic scale length $R$, it makes sense to order $s_s / n_s \Omega_s \sim \mathcal{O}(v_{sr} \delta / v_{s\theta}^{(0,0)})$. This will end up being $\mathcal{O}(\delta^2 \epsilon)$. 

To zeroth order in $\delta$ and $\epsilon$, Eq.~(\ref{eqn:vr}) requires that $v_{ir}$ and $v_{er}$ vanish and Eq.~(\ref{eqn:vtheta}) gives that the azimuthal motion is determined by the $\bE \times \bB$ drift:
\begin{align}
v_{i \theta}^{(0,0)} = v_{e \theta}^{(0,0)} = - \frac{E}{B} \,. 
\end{align}
The next order in $\delta$ includes the diamagnetic drift and an $\bF \times \bB$ drift associated with the centrifugal force and evaluated using $v_{s\theta}^{(0,0)}$:
\begin{gather}
v_{s \theta}^{(1,0)} = \frac{p_s'}{n_s m_s \Omega_s} - \frac{1}{\Omega_s} \frac{E^2}{r B^2} \, . \label{eqn:vtheta1}
\end{gather}
Eq.~(\ref{eqn:vtheta1}) can be used to get the leading-order nonvanishing radial velocities from the general expression in Eq.~(\ref{eqn:vr}): 
\begin{align}
v_{ir}^{(1,1)} = & \frac{T_i \nu_{ie}}{m_i \Omega_i^2} \bigg[ \bigg( - \frac{Z T_e}{T_i} \frac{p_e'}{p_e} - \frac{p_i'}{p_i} \bigg) \nonumber \\
&+ \bigg( 1 + \frac{Z m_e}{m_i} \bigg) \frac{m_i}{T_i} \frac{E^2}{r B^2} + \frac{3 Z T_e'}{2 T_i} \bigg] \label{eq:vr11}
\end{align}
and 
\begin{align}
v_{er}^{(1,1)} = v_{ir}^{(1,1)}. 
\end{align}
This velocity, together with the steady-state continuity equation, governs classical particle transport: 
At this order, there are radial drifts due to azimuthal friction, but they are ambipolar in the sense that they do not produce net radial current, only a net radial mass flow. 

To the next order in $\delta$, the radial motion is 
\begin{align}
v_{sr}^{(2,1)} &= - \frac{v_{sr}^{(1,1)}}{r \Omega_s} \frac{\partial \big( r v_{s \theta}^{(0,0)} \big)}{\partial r} - \frac{(\nabla \cdot \pi_s)_\theta}{n_s m_s \Omega_s} \nonumber \\
&\hspace{30 pt} + \frac{s_s^n (v_{s \theta}^\text{src} - v_{s \theta}^{(0,0)})}{n_s \Omega_s} + \frac{\nu_{ss'} \Delta v_\theta^{(2,0)}}{\Omega_s} \, . \label{eqn:vr3}
\end{align}
The first term on the RHS of Eq.~(\ref{eqn:vr3}) is the non-ambipolar radial motion resulting from the interplay between frictional and inertial effects; intuitively, it is the Coriolis $\bF \times \bB$ drift that results from the next-lowest order radial flow. The second term is the $\bF \times \bB$ from the azimuthal viscous force. The third is an effective $\bF \times \bB$ due to the momentum flux from particle sources and sinks. The last --- which will not produce net charge transport at this order --- is an ambipolar frictional flux resulting from higher-order corrections to $v_{s \theta}$, including those due to $(\nabla \cdot \pi_s)_r$. The expression $\Delta v_{\theta}^{(2,0)}$ is the difference between the ion and electron azimuthal flows of $\mathcal{O}(\delta^2)$. 

It is necessary to verify, since the relative size of $\epsilon$ and $\delta$ has not been specified, that the current is not sensitive to this ordering. 
$v_{sr}^{(k,0)}$ and $v_{sr}^{(0,k)}$ vanish for all $k$ since there are no radial flows that could compete with $v_{sr}^{(1,1)}$ in the inertial part of Eq.~(\ref{eqn:vr3}). There could conceivably be other orders contributing to the $\Delta v_{\theta}^{(2,0)}$ in Eq.~(\ref{eqn:vr3}), but the resulting flows do not contribute to the leading-order current or particle flux anyway. 

Moreover, no term other than $v_{sr}^{(2,1)}$ will produce the leading-order current. When constructing a non-ambipolar radial flow, one factor of $\delta$ is needed to produce any differential azimuthal motion and a second factor of $\delta$ is needed to make the drift due to $R_{s \theta}$ in Eq.~(\ref{eqn:vr}) not be intrinsically ambipolar. Without one factor of $\epsilon$, there is no friction force to begin with. As such, there is no need to worry about competition from, e.g., $\mathcal{O}(\delta^3)$ or $\mathcal{O}(\delta \epsilon^2)$ radial flows; the leading-order current comes from the motion described by Eq.~(\ref{eqn:vr3}). 

With that in mind, the current $j_r = n_e e (v_{ir}^{(2,1)} - v_{er}^{(2,1)})$ can be written as 
\begin{align}
j_{r} &= - \frac{n_i m_i v_{ir}^{(1,1)}}{r B} \frac{\partial \big( r v_{i \theta}^{(0,0)} \big)}{\partial r} - \frac{n_e m_e v_{er}^{(1,1)}}{r B} \frac{\partial \big( r v_{e \theta}^{(0,0)} \big)}{\partial r} \nonumber \\
&\hspace{10 pt} - \frac{(\nabla \cdot \pi_i)_\theta}{B} - \frac{(\nabla \cdot \pi_e)_\theta}{B} \nonumber \\
&\hspace{10 pt} + \frac{m_i s_i \big(v_{i \theta}^\text{src} - v_{i \theta}^{(0,0)}\big)}{B} + \frac{m_e s_e \big(v_{e \theta}^\text{src} - v_{e \theta}^{(0,0)} \big)}{B} \, . \label{eqn:jrArbitraryViscosityWithElectronMass}
\end{align}
So far, the generalization of this calculation to the case of multiple ion species would largely be trivial; the friction would have to include interactions between all pairs of species, and the thermal friction takes a slightly different form for particles of general mass, but otherwise the calculation would be the same. However, it is possible to simplify $j_r$ in the case of a single ion species by making use of the small mass ratio. Dropping anything of $\mathcal{O}(m_e/m_i)$, and defining $\pi \doteq \pi_i + \pi_e$, Eq.~(\ref{eqn:jrArbitraryViscosityWithElectronMass}) is 
\begin{align}
j_{r} &= - \frac{(\nabla \cdot \pi)_\theta}{B} +
\frac{Z e n_i T_i \nu_{ie}}{m_i \Omega_i^3} \frac{1}{r} \frac{\partial (r E / B)}{\partial r} \nonumber \\
&\hspace{50 pt} 
\times \bigg[ \frac{m_i}{T_i} \frac{E^2}{r B^2} + \frac{3ZT_e'}{2T_i} - \frac{Z T_e}{T_i} \frac{p_e'}{p_e} - \frac{p_i'}{p_i} \bigg] \nonumber \\
&\hspace{30 pt}+ \frac{m_i s_i (v_{i \theta}^\text{src} + E/B)}{B} \, . \label{eqn:jrArbitraryViscosity}
\end{align}
This can be separated out into several terms, with different scalings and behaviors, as we will discuss in the next section.

\section{Conservation Equations and Nondimensionalization} \label{sec:nondimensionalization}

In order to relate our sources of charge and current to the density, temperature, and rotation profiles in the system, we use the particle and charge conservation equations.
In a cylindrically symmetric system, in the limit where axial losses and gradients are small compared to their radial counterparts, these are given by:
\begin{gather}
s_i = \frac{1}{r} \frac{\partial}{\partial r} r n_i v_{ir}^{(1,0)}\\
e(Z s_i - s_e) = \frac{1}{r} \frac{\partial}{\partial r} r j_r.
\end{gather}
Integrating, we obtain
\begin{align}
S_i(r) &= r n_i v_{ir}^{(1,1)} \label{eqn:basicContinuity}\\
C(r) &= r j_r, \label{eqn:basicChargeConservation}
\end{align}
where we have defined the integrated source functions
\begin{align}
S_s(r) &= \int_0^r s_s r' dr'\\
C(r) &= e \int_0^r (Z s_i - s_e) r' dr'. \label{eq:norm2}
\end{align}

In order to understand the relationships between the different current terms, it is useful to introduce a number of dimensionless parameters. To that end, let $R$ be the radial device size, $B_0$ be the on-axis field strength, and $n_0$, $T_0$, $\eta_{i10}$ and  be a characteristic density, temperature, and Braginskii perpendicular viscosity (discussed in the next section) respectively. Define the frequencies 
\begin{align}
\Omega_{i0} &\doteq \frac{Z e B_0}{m_i} \label{eq:norm1}\\
\nu_{ie0} &\doteq \frac{Z^2 e^4 m_e^{1/2} n_0 \log \Lambda_{ie}}{3 (2 \pi)^{3/2} \epsilon_0^2 m_i T_0^{3/2}} \\
\nu_{ii0} &\doteq \frac{Z^4 e^4 n_0 \log \Lambda_{ii}}{12 \pi^{3/2}  \epsilon_0^2 m_i^{1/2} T_0^{\,3/2}} \\
\omega_{thi0} &\doteq \frac{v_{thi}}{r} = \frac{\sqrt{T_0 / m_i}}{R} \\
\omega_\text{rot} &\doteq \frac{v_{s \theta}^{(0,0)}}{r} = - \frac{E}{r B} \, .
\end{align}
Here $\log \Lambda_{ie}$ and $\log \Lambda_{ii}$ are Coulomb logarithms.
Note that $\omega_\text{rot}$ is the only frequency here with radial dependence, as it represents a dynamical variable which will be solved for.

Finally, define the total integrated source rate for particles and charge $S_0 = S_i(R)$, and $C_0 = C(R)$.
In terms of these constants, the \emph{total} number of ions and charge per unit axial length added to or extracted from the system per unit time are $2\pi S_0$ and $2\pi C_0$ respectively.

\begin{table}
	\begin{center}
		\begin{tabular}{| c | c |}
			\hline
			Parameter & Definition \\
			\hline
			$\tr$ & $r/R$ \\
			\hline
			$\tB$ & $B/B_0$ \\
			$\tT_s$ & $T_s / T_0$ \\
			$\tilde{S}$ & $S_i(r) \big/ S_0$ \\
			$\tilde{C}$ & $C(r) \big/ C_0$ \\
			\hline
			$\tn_s$ & $n_s / n_0$ \\
			$\tilde{\omega}_\text{rot}$ & $\omega_\text{rot} \big/ \omega_{thi0}$ \\
			$\tilde{\eta}$ & $\eta_{i1} / \eta_{i10}$ \\
			\hline
		\end{tabular}
	\end{center}
	\caption{Normalized variables used throughout the paper, based on quantities in Eqs.~(\ref{eq:norm1}-\ref{eq:norm2}), as well as Eq.~(\ref{eqn:etai1}). Horizontal lines delineate coordinate variables, specified profiles, and finally dynamical variables which are solved for. The viscosity $\tilde{\eta}$ falls in the final category due to its density-dependence.} \label{table:dimensionlessVariables}
\end{table}

\begin{table}
	\begin{center}
		\begin{tabular}{| c | c | c | }
			\hline
			Parameter & Definition & Interpretation\\
			\hline
			$\mathcal{P}$ & $S_0 \big/ (N_0 \nu_{ie0})$ & Particle Outflow\\
			$\mathcal{I}$ & $C_0 / \left[A (\rho_{i0}/R) (Z e N_0 \nu_{ii0}) \right]$ & Current\\
			$\mathcal{Q}$ & $A^{-1} \nu_{ie0}/\nu_{ii0}$ & Inertia / Viscosity\\
			\hline
		\end{tabular}
	\end{center}
	\caption{Dimensionless parameters which characterize the solution. Here, $N_0 = n_0 \rho_{i0}^2$ is the number of ions per unit length in a cylinder of area $\rho_{i0}^2$; this quantity appears repeatedly as an important characteristic of the system. $A$ is a numerical constant, $A = 3/10\sqrt{2}$.} \label{table:dimensionlessParameters}
\end{table}

This collection of parameters can be used to construct the normalized variables listed in Table~\ref{table:dimensionlessVariables}. 
These variables are characteristically $\mathcal{O}(1)$ by construction (at least in the case of an approximately sonic flow, for $\tilde{\omega}_\text{rot}$). 

Substitution of these variables into the continuity Eq.~(\ref{eqn:basicContinuity}) and charge conservation Eq.~(\ref{eqn:basicChargeConservation}), using the velocity in Eq.(\ref{eq:vr11}) and the current in Eq.~(\ref{eqn:jrArbitraryViscosity}), as well as the Braginskii transport coefficients,\cite{Braginskii1965} yields several dimensionless constants, listed in Table \ref{table:dimensionlessParameters}. 
These parameters determine the characteristic sizes of the relevant particle fluxes and currents in the systems, as a function of the typical gradients.
They depend on the relation between the source terms $S_0$ (for particles) and $C_0$ (for currents), on the number of particles $N_0$ in a cylinder of area $\rho_{i0}^2$, and on the relevant collision frequencies.

In terms of these parameters, our nondimensionalized continuity equation becomes:
\begin{align}
\mathcal{P} \, \tilde{S} = & \, \tr \frac{\tn_i \tn_e \tT_i }{\tT_e^{\,3/2} \tB^2} \bigg[ \bigg( - \frac{Z \tT_e}{\tT_i} \frac{\partial_{\tr} \tilde{p}_e}{\tilde{p}_e} - \frac{\partial_{\tr} \tilde{p}_i}{\tilde{p}_i} \bigg) \nonumber \\
&\hspace{30 pt}+ \bigg( 1 + \frac{Z m_e}{m_i} \bigg) \tr \, \tilde{\omega}_\text{rot}^2 + \frac{3 Z \partial_{\tr} \tT_e}{2 \tT_i} \bigg] . \label{eqn:dimensionlessContinuity}
\end{align}
Here, the RHS represents the charge added to the system, whereas the LHS determines the response of the system to the added charge. 
Because all tilde quantities (other than $\tilde{\omega}_\text{rot}$) are $\mathcal{O}(1)$, the dimensionless constant $\mathcal{P}$ determines the strength of the system response.
If the normalizations are chosen appropriately, if the flow is not substantially supersonic, and if the gradient scale lengths of the various parameters are not small compared to $R$, then Eq.~(\ref{eqn:dimensionlessContinuity}) implies that $\mathcal{P}$ is not larger than $\mathcal{O}(1)$. 
This means that we cannot drive fluxes that are too large through our system without creating extremely large gradients, which will likely lead to instabilities and anomalous transport.
Thus, $\mathcal{P} \lesssim \mathcal{O}(1)$ mathematically formalizes the requirement for smooth, stable solutions with device-scale gradients.

Current conservation takes a similar form:
\begin{align}
\mathcal{I} \tilde{C} &= \tr \left(\mathcal{J}^\text{viscous} + \mathcal{J}^\text{inertial} + \mathcal{J}^\text{src}  \right). \label{eqn:dimensionlessChargeConservation}
\end{align}
We will explore each of these terms in detail; the final result is in Eq.~(\ref{eq:Ifinal}).

\section{Physical Mechanisms of Cross-Field Current} \label{sec:mechanisms}

The various terms in Eq.~(\ref{eqn:dimensionlessChargeConservation}) describe cross-field currents that come from three distinct physical mechanisms. The following subsections discuss their physical origins and relative sizes. 

\subsection{Viscous Current} \label{subsec:viscousCurrent}
\begin{figure}
	\centering
	\begin{tikzpicture} 
	
	\draw[thick, blue, dashed] (0,4) arc (90:180:4);
	\draw[thick, blue, dashed, opacity=0.4] (0,3.7) arc (90:180:3.7);
	\draw[thick, blue, dashed, opacity=0.4] (0,4.3) arc (90:180:4.3);
	
	\draw[->, very thick, blue] (-4,0) -- (-4,1.2);
	\draw[->, very thick, blue, opacity=0.4] (-3.7,0) -- (-3.7,1.0);
	\draw[->, very thick, blue, opacity=0.4] (-4.3,0) -- (-4.3,1.4);
	\node at (-2.2,.5) {1. Sheared $\bv_a^{(0,0)}/r$};
	
	\draw[->, very thick, black] ({4*cos(120)},{4*sin(120)}) -- ({4*cos(120)-1.2*sin(120)},{4*sin(120)+1.2*cos(120)});
	\draw[->, very thick, blue] ({4*cos(120)},{4*sin(120)}) -- ({5.2*cos(120)},{5.2*sin(120)});
	\node at (-4.3, 3.0) {2. $(\nabla \cdot \pi_a)_\theta \, \hat \theta$};
	I
	\node at (-4.7, 4.4) {3. $\bv^{(2,1)}_{a,\text{visc}} \propto (\nabla \cdot \pi_a)_\theta \, \hat \theta \times \bB$};
	\draw[very thick, black] (-.5,1) circle (.2);
	\draw[black, fill] (-.5,1) circle (.05);
	\node at (0,1) {$\bB$};
	
	\draw[->, very thick, black] ({-.5+.2*cos(135)},{1+.2*sin(135)}) -- ({-.5+.6*cos(135)},{1+.6*sin(135)});
	\node at (-.6,1.6) {$\bE$};
	
	\end{tikzpicture}
	\caption{This schematic illustrates the physics of the viscous current. Shear in the leading-order azimuthal flow (1) leads to a viscous force with a component in the azimuthal direction (2), whose sign will depend on $\bv_a(r)$ and $\eta_{i1}(r)$. The resulting $\mathcal{O}(\delta^2 \epsilon)$ $\mathbf{F} \times \bB$ flow carries a current (3).} \label{fig:viscousCartoon}
\end{figure}
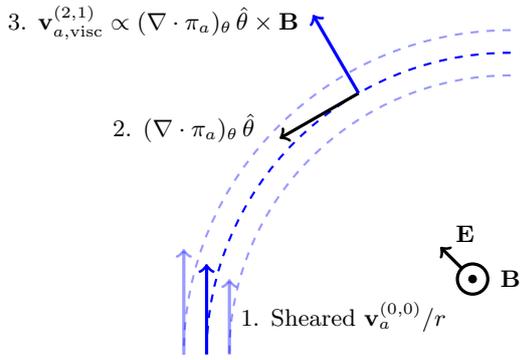

The viscous current is a radial $\mathbf{F} \times \bB$ drift resulting from the azimuthal viscous force (see Figure~\ref{fig:viscousCartoon}): 
\begin{gather}
j_r^\text{viscous} = - \frac{(\nabla \cdot \pi)_\theta}{B} \, .
\end{gather}
If $\pi = \sum_s \pi_s$ is the Braginskii viscosity,\cite{Braginskii1965} then keeping the contributions from the leading-order flow $\bv_s^{(0,0)} = v_{s \theta}^{(0,0)}(r) \hat \theta$, 
\begin{align}
(\nabla \cdot \pi_s)_\theta &= - \frac{1}{r^2} \frac{\partial}{\partial r} \bigg[ r^3 \eta_{s1} \frac{\partial}{\partial r} \bigg( \frac{v_{s \theta}^{(0,0)}}{r} \bigg) \bigg] \label{eqn:viscosity} \\ 
(\nabla \cdot \pi_s)_r &= - \frac{1}{r^2} \frac{\partial}{\partial r} \bigg[ r^3 \eta_{s3} \frac{\partial}{\partial r} \bigg( \frac{v_{s \theta}^{(0,0)}}{r} \bigg) \bigg] \label{eqn:gyroviscosity} 
\end{align}
where the relevant viscosity coefficients are 
\begin{align}
\eta_{i1} &= \frac{3 \nu_{ii} p_i}{10 \sqrt{2} \Omega_i^2} & \label{eqn:etai1}
\eta_{e1} &= \frac{0.51 \nu_{ee} p_e}{\sqrt{2} \Omega_e^2}\\
\eta_{i3} &= \frac{p_i}{2 \Omega_i} &
\eta_{e3} &= \frac{p_e}{2 \Omega_e} \, .
\end{align}
The mass dependence in these coefficients means that it is generally safe to set $\pi = \pi_i$. 

The azimuthal viscous force results in a radial $\bF \times \bB$ drift, which carries ion current.
This current has the behavior one would intuitively expect from a viscosity: it acts to relax shear in the $\bE \times \bB$ flow. 
It is somewhat suppressed because it is a collisional finite-Larmor-radius effect. 
The radial viscous force comes from the gyroviscosity. Its form is very similar to that of its azimuthal counterpart, but (perhaps counterintuitively) it is substantially larger --- in the case of the ions, by a factor of $10 \Omega_i / 3 \sqrt{2} \nu_{ii} \sim 2/\epsilon$. Its contribution to $j_r$ is nonetheless small compared to the contribution of the azimuthal viscous force, since the current associated with this large azimuthal effect is ordered down by a factor of $\mathcal{O}(\epsilon \delta)$ from the original azimuthal drift, as discussed in the next section.
In other words, $(\nabla \cdot \pi)_\theta$ is more important for conductivity calculations because its $\bF \times \bB$ flow is non-ambipolar to start with, whereas $(\nabla \cdot \pi)_r$ can only drive current indirectly through inertial effects. 

In terms of dimensionless variables, the viscous current can be written as 
\begin{align}
\mathcal{J}^\text{viscous} &= \frac{1}{\tilde{r}^2 \tilde{B}} \frac{\partial}{\partial \tilde{r}} \bigg[ \tilde{r}^3 \tilde{\eta} \frac{\partial \tilde{\omega}_\text{rot}}{\partial \tilde{r}} \bigg] . 
\end{align}
When using the Braginskii viscosity, define $\tilde{\eta} \doteq \tn^2 / \tT^{1/2} \tB^2$ and $\eta_{i10} \doteq \eta_{i1} / \tilde{\eta}$. 
There are two classes of rotation profile that make the viscous current vanish. The first consists of any solid-body rotation profile. This is straightforward; without shear, the viscous force vanishes. The second class of profiles can be defined by 
\begin{gather}
\frac{\partial \tilde{\omega}_\text{rot}}{\partial \tr} = \frac{D}{\tr^3 \tilde{\eta}} \label{eqn:irrotationalOmega}
\end{gather}
for any constant $D$. These profiles are closely related to the irrotational vortices often found in neutral fluids. The intuition behind Eq.~(\ref{eqn:irrotationalOmega}) is much the same: there is viscous stress everywhere, but that stress produces no net force on any particular fluid element away from the origin, where these profiles are generally not physical (see, e.g., Kundu, Cohen, and Dowling\cite{KunduCohenDowling}).

\begin{figure}
\centering
\begin{tikzpicture} 

\draw[thick, blue, dashed] (0,4) arc (90:180:4);
\draw[thick, red,  dashed] (0,3.7) arc (90:180:3.7);

\draw[->, very thick, blue] (-4,0) -- (-4,1.5);
\draw[->, very thick,  red] (-3.7,0) -- (-3.7,1.2);
\node at (-2.3,.5) {1. $\bv_a^{(1,0)} \neq \bv_b^{(1,0)}$};

\draw[->, very thick, black] ({4*cos(130)},{4*sin(130)}) -- ({4*cos(130)-1.2*sin(130)},{4*sin(130)+1.2*cos(130)});
\draw[->, very thick, blue] ({4*cos(130)},{4*sin(130)}) -- ({5.2*cos(130)},{5.2*sin(130)});
\node at (-4, 2.7) {2. $\mathbf{R}_{ab}$};
\node at (-4.2, 4.3) {3. $\bv^{(1,1)}_a \propto \mathbf{R}_{ab} \times \bB$};

\draw[->, very thick, black] ({4*cos(100)},{4*sin(100)}) -- ({4*cos(100)-.6*sin(100)},{4*sin(100)+.6*cos(100)});
\draw[->, very thick, blue] ({4*cos(100)},({4*sin(100)}) -- ({4.8*cos(100)},{4.8*sin(100)});
\node at (-1.7, 4.3) {4. $\mathbf{F}_\text{cor}$};
\node at (-1.5, 5.1) {5. $\bv^{(2,1)}_{a,\text{iner}} \propto \mathbf{F}_\text{cor} \times \bB$};

\draw[very thick, black] (-.5,1) circle (.2);
\draw[black, fill] (-.5,1) circle (.05);
\node at (0,1) {$\bB$};

\draw[->, very thick, black] ({-.5+.2*cos(135)},{1+.2*sin(135)}) -- ({-.5+.6*cos(135)},{1+.6*sin(135)});
\node at (-.6,1.6) {$\bE$};

\end{tikzpicture}
\caption{This figure shows the physics behind the inertial current. Differences in local velocities $v_{s \theta}^{(1,0)}$ between different species (1) lead to azimuthal friction forces (2). These produce ambipolar radial drifts $v_{sr}^{(1,1)}$ (3) which in turn produce azimuthal Coriolis forces (4). The resulting Coriolis $\mathbf{F} \times \bB$ drifts $v_{sr}^{(2,1)}$ (5) carry the current.} \label{fig:inertialCartoon}
\end{figure}
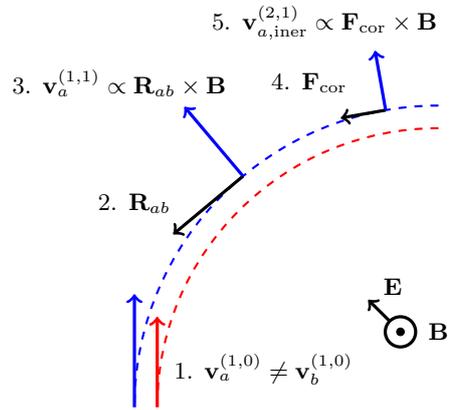

\subsection{Inertial Current} \label{subsec:inertialCurrent}
The inertial current can be written as 
\begin{align}
j_{r}^\text{inertial} &= \frac{Z e n_i T_i \nu_{ie}}{m_i \Omega_i^3} \frac{1}{r} \frac{\partial (r E / B)}{\partial r} \nonumber \\
&\hspace{20 pt} \times \bigg[ \frac{m_i}{T_i} \frac{E^2}{r B^2} + \frac{3ZT_e'}{2T_i} - \frac{Z T_e}{T_i} \frac{p_e'}{p_e} - \frac{p_i'}{p_i} \bigg]. 
\end{align}
Physically, this current can be explained in terms of a series of $\mathbf{F} \times \bB$ drifts. Temperature gradients and differences in the centrifugal and diamagnetic drifts between species ($v_{\theta s}^{(1,0)}$, Eq.~(\ref{eqn:vtheta1})) produce a $\hat \theta$-directed friction between species. This friction force leads to a radial $\bF \times \bB$ drift ($v_{r s}^{(1,1)}$, Eq.~(\ref{eq:vr11})). This initial radial flow does not carry net current, but the motion results in an azimuthal Coriolis force, which depends on mass and therefore affects the ions far more than the electrons. The $\mathbf{F} \times \bB$ associated with this Coriolis force is the flow which carries the leading-order radial current. This sequence of effects is shown in Figure~\ref{fig:inertialCartoon}. 

In the limit where $\omega_\text{rot}$ is constant and where particle injection effects can be neglected, the inertial current is the only current. If, in addition, the densities and temperatures are constant, the current can be written as 
\begin{gather}
j_{r}^\text{inertial} = \frac{2 Z e n_i \nu_{ie}}{r^2 \Omega_i^3} \frac{E^3}{B^3} \, .
\end{gather}
This expression is equivalent to the nonlinear Ohm's law described by Rax \textit{et al.} \cite{Rax2019}

In terms of dimensionless variables, the inertial current can be written 
\begin{align}
\mathcal{J}^\text{inertial} &= - \frac{\mathcal{P} \, \mathcal{Q} \, \tilde{S}}{\tB \tr^2} \, \frac{\partial (\tr^2 \tilde{\omega}_\text{rot})}{\partial \tr} \, .
\end{align}
The inertial current vanishes wherever $\tilde{\omega}_\text{rot} \propto 1/\tr^2$. The Coriolis force can be understood as a mechanism that enforces the conservation of angular momentum during radial motion; these profiles have flows with uniform angular momentum, so it makes sense to find that they do not have Coriolis $\mathbf{F} \times \bB$ drifts. 

\subsection{Particle Injection Current} \label{subsec:injectionCurrent}
In a region where there are particle sources or sinks, there is an additional current: 
\begin{align}
j_r^\text{src} &= \frac{m_i s_i (v_{i \theta}^\text{src} + E/B)}{B} . \label{eqn:jInjection}
\end{align}
$v_{i \theta}^\text{src}$ term comes from the momentum injection due to sources and sinks and the $E/B$ term is inertial in the sense that it comes from the left-hand side of Eqs.~(\ref{eqn:ionMomentum}) and (\ref{eqn:electronMomentum}). 

However, this current can be explained more intuitively in terms of single-particle dynamics.\cite{Goertz1980, Rozhanskii1990, Rozhansky2008} Consider a particle placed in crossed $\bE = E \hat x$ and $\bB = B \hat z$ fields with some initial velocity $\bv_0 = v_{x0} \hat{x} + v_{y0} \hat{y}$ and position $\mathbf{r}_0 = x_0 \hat{x} + y_0 \hat{y}$. In the limit of locally constant fields, it will gyrate about an $x$-coordinate that is shifted from its initial position:
\begin{gather}
x_\text{gyro} = x_0 + \frac{m}{qB} \bigg( v_{y0} + \frac{E}{B} \bigg). \label{eqn:xGyro}
\end{gather}
Thus, there will be a current resulting from the fact that particles move from their initial to their average position as they are added to the system, which results in Eq.~(\ref{eqn:jInjection}).
When $v_{i \theta}^\text{src} = 0$, this has been termed the ``pickup current'' \cite{Goertz1980} or ``mass-loading current''.\cite{Rozhanskii1990, Rozhansky2008}

This current can be written in terms of dimensionless parameters as 
\begin{align}
\mathcal{J}^\text{src} &= \mathcal{Q} \mathcal{P} \, \frac{\partial \tilde{S}}{\partial \tilde{r}} \frac{(\tilde{\omega}_\text{src} - \tilde{\omega}_\text{rot})}{\tilde{B}} \, .
\end{align}
The dependence on $\partial \tilde{S} / \partial \tr$ follows from the reliance of this mechanism on the local value of $s_i$.

\subsection{Comparing the Currents}
Substituting in the expressions for the components of $\mathcal{J}$, Eq.~(\ref{eqn:dimensionlessChargeConservation}) becomes:
\begin{align}
\mathcal{I} \tilde{C} &= \frac{1}{\tilde{r} \tilde{B}} \frac{\partial}{\partial \tilde{r}} \bigg[ \tilde{r}^3 \tilde{\eta} \, \frac{\partial \tilde{\omega}_\text{rot}}{\partial \tilde{r}} \bigg] \nonumber \\
& \quad - \mathcal{Q} \mathcal{P} \bigg[ \frac{\tilde{S}}{\tilde{B} \tilde{r}} \frac{\partial (\tilde{r}^2 \tilde{\omega}_\text{rot})}{\partial \tilde{r}} + \tilde{r} \frac{\partial \tilde{S}}{\partial \tilde{r}} \frac{(\tilde{\omega}_\text{rot} - \tilde{\omega}_\text{src})}{\tilde{B}} \bigg]   . \label{eq:Ifinal}
\end{align}
$\mathcal{Q}$ is 
\begin{gather}
\mathcal{Q} = \frac{10 \sqrt{2} \nu_{ie0}}{3 \nu_{ii0}}
\end{gather}
and $\mathcal{I}$ is 
\begin{gather}
\mathcal{I} = \frac{10 \sqrt{2}}{3}\frac{C_0}{(\rho_{i0}/R) (Z e N_0 \nu_{ii0}) } \, . \label{eqn:currentComparison}
\end{gather}
Eq.~(\ref{eqn:currentComparison}) gives some insight into the relative sizes of the different currents. Depending on, for instance, the shape of $\tilde{\omega}_\text{rot}$, any of the mechanisms can be dominant. However, suppose that all gradient scale lengths are on the order of $R$. Recall that, in many cases, $\mathcal{P} \sim \mathcal{O}(1)$. When the currents are written in terms of dimensionless $\mathcal{O}(1)$ quantities, the inertial and particle injection currents are attached to a prefactor of $\mathcal{Q}$, which is small compared to 1. If $\mathcal{I}$ is also small compared to 1, then any current may dominate. If $\mathcal{I} \sim \mathcal{O}(1)$, then the viscous current likely dominates. If $\mathcal{I}$ is large compared to 1, then the aforementioned assumption about the gradient scale lengths is probably invalid. 

We can estimate the total current drawn by the machine by balancing the two sides of Eq.~(\ref{eqn:currentComparison}), yielding $\mathcal{I} \sim \tilde{\omega}_\text{rot}$, and solving for $2\pi C_0$.
This gives
\begin{align}
2 \pi C_0 &\sim \frac{3\pi }{5 \sqrt{2}} \left(\frac{\rho_{i0}}{R}\right) Z e N_0 \nu_{ii0} \tilde{\omega}_\text{rot}
\end{align}
so
\begin{align}
\frac{2 \pi C_0}{11 \text{ A/m}} & \sim Z^{2} \bigg( \frac{R}{10 \text{ cm}} \bigg)^{-1} \bigg( \frac{m_i}{m_p} \bigg) \bigg( \frac{B}{1 \text{ kG}} \bigg)^{-3} \notag\\
&\qquad \times \bigg( \frac{n_0}{10^{14} \text{ cm}^{-3}} \bigg)^{2} \bigg( \frac{\log \Lambda_{ii}}{10} \bigg) \tilde{\omega}_\text{rot}. 
\end{align}

\section{Calculating Rotation Profiles from Ohm's Law}  \label{sec:profile}

In a system with defined sources and sinks of charged particles, the steady-state total current through the system can often be written easily in terms of those sources and sinks. That information, combined with the appropriate Ohm's law, can be used to extract other useful information about the system. 

Suppose the particle sources and sinks are known, so that $S_s(r)$ and $C(r) = Z e S_i(r) - e S_e(r)$ are specified. 
Suppose the particle sources have $v_{i \theta}^\text{src} = 0$. Invoking Ohm's law and the continuity equations and simplifying, \begin{align}
\tr^3 \tilde{\eta} \, \tilde{\omega}_\text{rot}' - \mathcal{Q} \mathcal{P} \, \tr^2 \tilde{S} \, \tilde{\omega}_\text{rot} = \mathcal{I} \int_0^{\tr} \tr' \tilde{C}(\tr') \tilde{B}(\tr') d \tr'. \label{eqn:omegaODE}
\end{align}
This has a general solution for $\omega_\text{rot}$: 
\begin{align}
&\tilde{\omega}_\text{rot}(\tilde{r}) 
= e^{\mathcal{Q} \mathcal{P} \int_1^{\tilde{r}} (\tilde{S} / \tilde{x} \tilde{\eta}) d \tilde{x}} \bigg[ \tilde{\omega}_\text{rot}\big|_{\tr=1} \nonumber \\
&\hspace{0 pt}+ \mathcal{I} \int_1^{\tilde{r}} d\tilde{s} \bigg( \frac{e^{- \mathcal{Q} \mathcal{P} \int_1^{\tilde{s}} (\tilde{S} / \tilde{y} \tilde{\eta}) d\tilde{y}}}{\tilde{s}^3 \tilde{\eta}(\tilde{s})} \int_0^{\tilde{s}} \tilde{r}' \tilde{C}(\tilde{r}') \tilde{B}(\tilde{r}') d \tilde{r}' \bigg) \bigg] . \label{eqn:omegaGeneral}
\end{align}
for a boundary condition imposed at $R$.

For example, suppose $B$ were constant in space and $\omega(R) = 0$. For some $\alpha \geq 0$, suppose $C = C_0 (r/R)^\alpha$ and $S_i = S_{0} (r/R)^\alpha$. Then 
\begin{align}
\tilde{\omega}_\text{rot} &= - \frac{\mathcal{I}}{\mathcal{Q} \mathcal{P}} \frac{\tilde{B}}{\alpha+2} \nonumber \\
&\hspace{20 pt} \times \bigg[ 1 - \exp \bigg( - \mathcal{Q} \mathcal{P} \int_{\tilde{r}}^1 \frac{\tilde{s}^{\alpha-1} d\tilde{s}}{\tilde{\eta}(\tilde{s})} \bigg) \bigg]. 
\end{align}
The case of uniform volumetric particle injection is given by $\alpha = 2$. The case in which all particles are injected at $r = 0$ is given by $\alpha = 0$. Note that 
\begin{gather}
\tilde{\omega}_\text{rot} \rightarrow \begin{cases}
- \mathcal{I} \, \frac{\tilde{B}}{\alpha+2} \int_{\tr}^1 \frac{\tilde{s}^{\alpha-1} d \tilde{s}}{\tilde{\eta}(\tilde{s})} & \quad \mathcal{Q}\mathcal{P} \ll 1 \\
- \frac{\mathcal{I}}{\mathcal{Q} \mathcal{P}} \frac{\tilde{B}}{\alpha+2} & \quad \mathcal{Q}\mathcal{P} \gg 1 ,
\end{cases}
\end{gather}
consistent with the intuition from Section~\ref{sec:nondimensionalization}. 


Now consider a scenario in which $B$, $\eta_{i1}$, and $C$ have arbitrary shapes but where $S_i$ is not much larger than $Z_i S_i - S_e$. In this case, the orderings described in Section~\ref{sec:derivation} come out slightly differently; if the ambipolar flow is not much larger than the non-ambipolar flow, then the inertial current will be small compared to the viscous current. For example, if the system has a radial flow of electrons but no net flow of ions, then (using the same boundary condition as before) 
\begin{gather}
\tilde{\omega}_\text{rot}' = \frac{\mathcal{I}}{\tr^3 \tilde{\eta}} \int_0^{\tr} \tr' \tilde{C}(\tr') \tilde{B}(\tr') d \tr'
\end{gather}
and 
\begin{gather}
\tilde{\omega}_\text{rot} = - \mathcal{I} \int_{\tr}^1 \frac{d \tilde{s}}{\tilde{s}^3 \tilde{\eta}(\tilde{s})} \int_0^{\tilde{s}} \tr' \tilde{C}(\tilde{r}') \tilde{B}(\tilde{r}') d \tr' . 
\end{gather}

\begin{figure*}
\centering
\includegraphics[width=\linewidth]{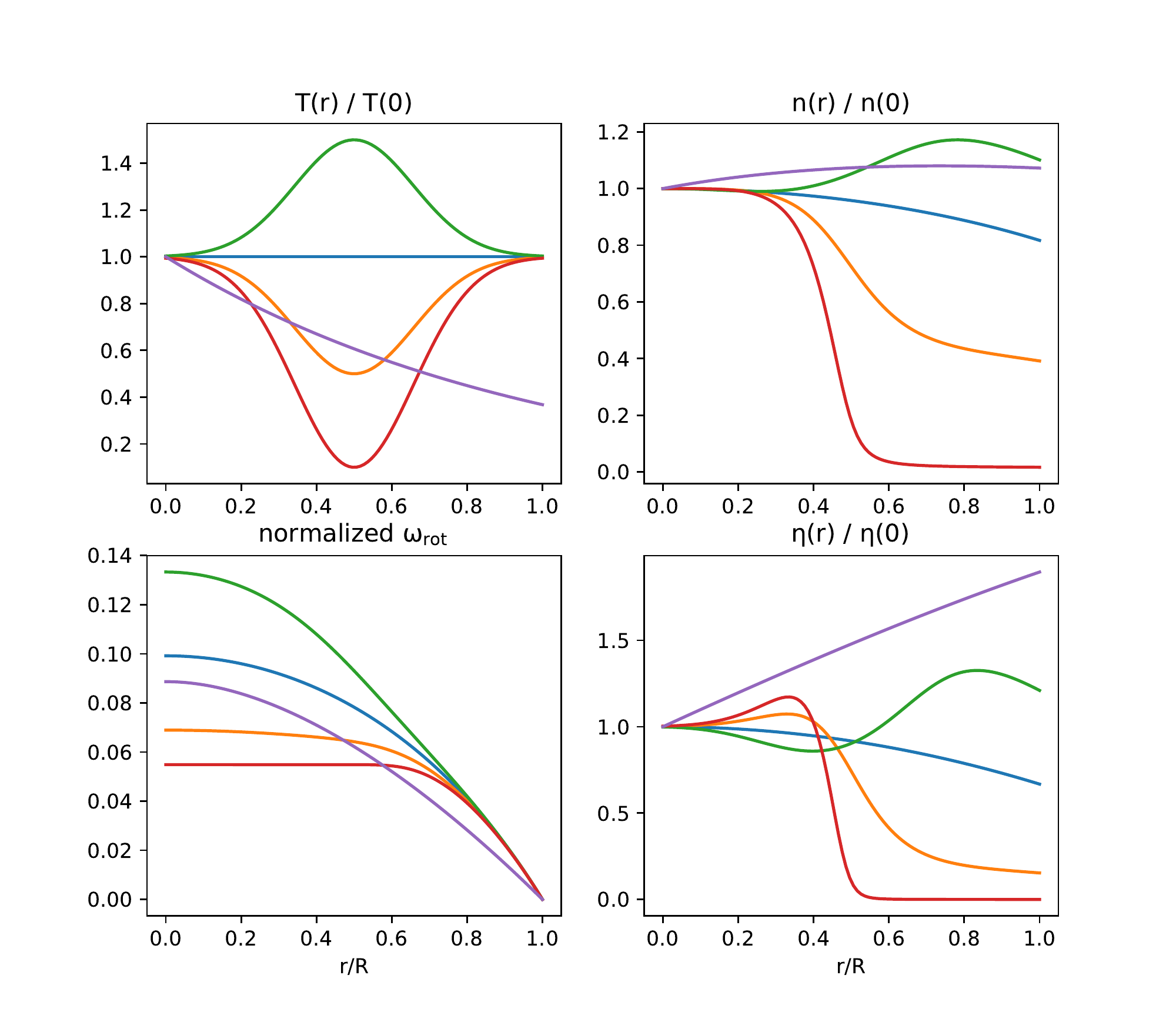}
\caption{This figure shows several profiles of $T$, $n$, $\omega_\text{rot}$, and $\eta_{i1}$ that are consistent with the particle and charge transport equations; here $T(r)$ was specified and the other consistent profiles were computed numerically. Curves of the same color correspond to the same scenario. }
\label{fig:profiles}
\end{figure*}

There are practical contexts in which it is important to control the rotation profile $\omega_\text{rot}(r)$. For example, the double well mass filter concept relies on a rotating plasma with a particular shear profile.\cite{Gueroult2014} It may be possible to control shear with end electrodes, but the calculations in this section suggest an additional strategy: that the viscosity, temperature, and particle input profiles can be used to control $\omega_\text{rot}(r)$. Viscosity depends on the density and temperature profiles as $\eta_{i1} \propto n_i^2 \, T_i^{-1/2}$; there may be contexts in which these are easier to manipulate than the electric field itself. 

Suppose some sufficiently sophisticated heating system could be used to specify $T(r)$. Then Eqs.~(\ref{eqn:dimensionlessContinuity}), (\ref{eqn:etai1}), and (\ref{eqn:omegaGeneral}) determine $n(r)$, $\eta_{i1}(r)$, and $\omega_\text{rot}(r)$. Figure~\ref{fig:profiles} shows several numerical solutions in the case of uniform particle and charge injection ($\tilde{S} = \tr^2$ and $\tilde{C} = - \tr^2$). If a particular $\omega_\text{rot}(r)$ were desirable, it might be possible to tailor $T(r)$ to get that profile. 

However, there are limits to what can be done with $T(r)$ alone. To see this, consider the following expression, which follows from Eq.~(\ref{eqn:omegaODE}):
\begin{gather}
\tilde{\eta} = \frac{\mathcal{Q}\mathcal{P} \tilde{S} \, \tilde{\omega}_\text{rot}}{\tr \tilde{\omega}_\text{rot}'} + \frac{\mathcal{I}}{\tr^3 \tilde{\omega}_\text{rot}'} \int_0^{\tr} \tr' \tilde{C}(\tr') \tilde{B}(\tr') d\tr' . 
\end{gather}
For instance, if $\tilde{\omega}_\text{rot}'$ is to change sign at $\tr$ without $\tilde{\eta}$ becoming singular, $\tilde{\omega}_\text{rot}(\tr)$ must be equal to $-(\mathcal{I} / \mathcal{Q} \mathcal{P} )(\int_0^{\tr} \tr' \tilde{C} \tilde{B} d\tr') / (\tr^2 \tilde{S})$. 
In fact, for the simple volumetric sources used in Figure~\ref{fig:profiles}, it is possible to show analytically that all solutions for $\omega_\text{rot}$ must be monotonic. 
Nonetheless, even for this very simple choice of sources, an experimentalist with good control over $T(r)$ could achieve a wide range of density and rotation profiles. 

\section{Discussion and Summary} \label{sec:discussion}

We described the motion of charge across magnetic field lines in a sheared rotating system and offered physical descriptions of the key terms in our derivation. We also suggested that sufficient control over the temperature and density profiles should allow control over the rotation profile $\omega_\text{rot}(r)$. 

Note that the form of Ohm's law presented here can be related to expressions in a number of places elsewhere in the literature. 
Rax \textit{et al.} \cite{Rax2019} analyzes a case which does not include shear. Their expression for Ohm's law includes an inertial current consistent with the one derived here, but because they studied solid-body rotation they do not include the viscous current. 
The review by Rozhansky \cite{Rozhansky2008} includes expressions for the viscous and mass-loading currents that agree with the ones derived here. He discusses inertial currents in general, but does not derive the Coriolis-driven inertial current that appears in a rotating plasma. 
The current edition of the book by Helander and Sigmar \cite{HelanderSigmar} includes a nice derivation of the inertial current, but their treatment of the viscosity (in the case of an axisymmetric rotating plasma) does not agree with the one presented here, essentially because of a mismatch between Eq.~(\ref{eqn:viscosity}) and their Eq.~(5.12). The derivation of Eq.~(\ref{eqn:viscosity}) is given in the Appendix. 

Other treatments of this problem have generally not considered the gyroviscosity, which produces $(\nabla \cdot \pi)_r$ as per Eq.~(\ref{eqn:gyroviscosity}). In many situations, this is not a major problem. After all, for the ordering used in this paper, $(\nabla \cdot \pi)_r$ does not contribute to the leading-order current despite the fact that it is substantially larger than $(\nabla \cdot \pi)_\theta$. Nonetheless, there is some value in being aware of this term. For instance, for a plasma approaching the Brillouin limit, $\delta$ will no longer be small and the gyroviscous $(\nabla \cdot \pi)_r$ will no longer necessarily be suppressed relative to $\nabla p$ (though the form of the viscosity tensor might need to be revisited in that limit). 

This derivation was done for a quiescent plasma. However, some models for anomalous transport provide prescriptions for alternate forms of the viscous force density $\nabla \cdot \pi$. To the extent that the rest of the bulk behavior of the plasma is unchanged, it is possible to substitute alternate viscosities into results like Eq.~(\ref{eqn:jrArbitraryViscosity}) or Eq.~(\ref{eqn:omegaGeneral}) to understand how to control current flow or rotation profiles in turbulent regimes. 

A number of authors have written about models in which (in an appropriate regime) anomalous transport produces something that looks like cross-field classical transport but with a higher effective collision frequency.\cite{Artsimovich1971, Finn1992, Rognlien1999} Rognlien and Ryutov called this ``pseudoclassical transport".\cite{Rognlien1999} 

Finn, Guzdar, and Chernikov studied cross-field transport for the case of stochastic field lines.\cite{Finn1992} Their work suggested a viscosity coefficient 
\begin{gather}
\eta_{\perp \text{F}} = m_i n_i \bigg( \frac{\delta B}{B} \bigg)^2 \ell_c c_s. 
\end{gather}
Here $\delta B$ is the magnitude of the non-axisymmetric field fluctuations, $\ell_c$ is the turbulent correlation length (often the connection length), and $c_s$ is the sound speed. 
There are laboratory observations in which this viscosity model appears to accurately describe momentum transport.\cite{Almagri1998} In such a case, Eq.~(\ref{eqn:omegaGeneral}) describes how $\omega_\text{rot}(r)$ reacts to the suppression or increase of turbulence. 

There is no guarantee that an anomalous viscosity should respect the same ordering as the Braginskii viscosity, either for $(\nabla \cdot \pi)_r$ or $(\nabla \cdot \pi)_\theta$. If either of these is very large, then the expression for the inertial current might need to be reconsidered. 

There are a number of other contexts in which the cross-field conductivity described in this paper would have to be modified. 
Cross-field dynamics can turn out quite differently in plasmas with significant ion-neutral collisions, weak magnetization, or more complicated geometries.\cite{Bonnevier1966, Lehnert1973, HazeltineMeiss, Rozhansky2008, Meier2012, Rax2015, Ochs2017, Zweben2018, Poulos2019, Jin2019} 
In the regime where this calculation does apply, it suggests some novel techniques with which the rotation profile might be controlled. Neutral beams, pellet injection, and electron injection can help shape the rotation profile by changing $\mathcal{P}$ and $\mathcal{I}$. Moreover, charge transport is substantially temperature-dependent, so heating or cooling particular regions of the plasma can change $\omega_\text{rot}(r)$. These techniques might be used either in place of more conventional techniques for rotation control or as a supplement to them. 

\begin{acknowledgements}
This work was supported by NNSA 83228-10966 [Prime No. DOE (NNSA) DE-NA0003764], and by NSF PHY-1506122. 
One author (IEO) also acknowledges the support of the DOE Computational Science Graduate Fellowship (DOE grant number DE-FG02-97ER25308).
\end{acknowledgements}

\begin{appendix}

\section{The Braginskii Viscosity in a Rotating Plasma}		

This paper uses Eqs.~(\ref{eqn:viscosity}) and (\ref{eqn:gyroviscosity}) for the components of $\nabla \cdot \pi$ when $\bv = v(r) \hat \theta$. This appendix will present two approaches to calculating these components from Braginskii's viscosity expression,\cite{Braginskii1965} which was originally given in Cartesian coordinates. 

\subsection{Viscosity in Arbitrary Orthonormal Coordinates} \label{subsec:tensorial}

When we wish to include viscous effects while working with curvilinear coordinates, we encounter a problem: Braginskii's viscosity tensor (Eqs.~(4.41) and (4.42)) is not expressed in a coordinate-invariant way, although the later, low-flow-ordered transport equations relevant to tokamak physics included more adaptable expressions.\cite{Mikhailovskii1984, Catto2004, Simakov2004}
This part of the Appendix will express the original Braginskii viscosity in a tensorial way that is easy to adapt to many coordinate systems. 

We will start by reviewing the original form of the Braginskii viscosity matrix.
The proper covariant expression should then be the tensorial expression which reduces to this expression.
Finally, we will explain how our results provide an easy recipe to calculate the Braginskii viscous force in an arbitrary orthonormal coordinate system.

Our notation will follow, e.g., Carroll's book on general relativity.\cite{Carroll}
Thus the covariant derivative is denoted
\beq
\nabla_{\mu} V^{\nu} \doteq \partial_\mu V^{\nu} + \Gamma^{\nu}_{\mu \sigma} V^\sigma.
\eeq
Here, $\Gamma^{\nu}_{\mu \sigma}$ is the Christoffel symbol, given by
\beq
\Gamma^{\nu}_{\mu \sigma} \doteq \frac{1}{2} g^{\nu \lambda} \lp \partial_\mu g_{\sigma \lambda} + \partial_\sigma g_{\lambda \mu} - \partial_\lambda g_{\mu \sigma}\rp. \label{eq:Christoffel}
\eeq
In addition, we will adopt the convention of using $v^i$ to refer to the conventional (orthonormal-basis) velocity, and $u^i$ to refer to the generalized velocity, which is a proper (contravariant) vector in the appropriate coordinate basis.
This will be important in the $\theta$ direction in cylindrical coordinates, since
\begin{align}
u^\theta &= \frac{d \theta}{d t}\\
v^\theta &= r \frac{d \theta}{d t}.
\end{align}
Whether we write $v^i$ with an upper and lower index has no significance; however, an upper vs lower index on $u^i$ denotes a different object (vector vs. one-form).
For a diagonal metric, these quantities are related by
\begin{equation}
    \sqrt{g_{ii}} u^i = v^i = u_i/\sqrt{g_{ii}}.
\end{equation}
Similarly, we will adopt the convention of using $h^i$ to refer to the orthonormal magnetic field unit vector, and $b^i$ to refer to the contravariant vector.

The Braginskii viscosity matrix is given by:\cite{Braginskii1965}
\begin{align}
\pi_{ab} &= -\eta_0 \, {^{0}}W_{ab} - \eta_1 \, {^{1}}W_{ab} - \eta_2 \, {^{2}}W_{ab} \nonumber \\
&\hspace{90 pt}+ \eta_3 \, {^{3}}W_{ab} + \eta_4 \, {^{4}}W_{ab}, \label{eq:viscousSum}
\end{align}
where
\begin{align}
^0W_{ab} &= \frac{3}{2} \big(h_a h_b - \frac{1}{3} \delta_{ab} \big) \big(h_m h_n - \frac{1}{3} \delta_{mn}\big) W_{mn} \label{eqn:W0}\\
^1W_{ab} &= \delta^{\perp}_{am} W_{mn} \delta^{\perp}_{nb} + \frac{1}{2} h_m W_{mn} h_n \delta^{\perp}_{ab}
\\
^2W_{ab} &= \delta^{\perp}_{am} W_{mn} h_n h_b + h_a h_m W_{mn} \delta^{\perp}_{nb} \\
^3W_{ab} &= \frac{1}{2} \big( \delta^{\perp}_{am} W_{mn} \tilde{\epsilon}_{nbk} h_k - \tilde{\epsilon}_{amk} h_k W_{mn} \delta^{\perp}_{nb}  \big)\\
^4W_{ab} &= h_a h_m W_{mn} \tilde{\epsilon}_{nbk} h_k - \tilde{\epsilon}_{amk} h_k W_{mn} h_n h_b . 
\end{align}
Here, $h_i$ is the $i$th component of the magnetic field unit vector, $\delta_{ij}$ is the Kronecker delta, $\teps_{ijk}$ is the Levi-Civita symbol, $\delta^{\perp}_{ij} = \delta_{ij} - h_i h_j$, and $W_{ij}$ is Braginskii's traceless rate-of-strain tensor, given by
\beq
W_{ij} \equiv \pa{v^i}{x^j} + \pa{v^j}{x^i} - \frac{2}{3}\delta_{ij} \nabla \cdot \mathbf{v}. \label{eqn:W}
\eeq
In the above definitions, repeated indices are summed over, even if they do not appear in upper-lower pairs. 
(Note that, in contrast to Braginskii, we have moved the matrix label for the $^i\te{W}{_{ab}}$ to the left hand side of the symbol, to make it clear that this is not a tensor index, but rather a label to a specific tensor.)

Now we need to find the covariant generalizations of the various quantities that appear in Eqs.~(\ref{eqn:W0}-\ref{eqn:W}). Make the substitutions 
\begin{gather}
\delta_{ij} \Longrightarrow g_{ij} \\
\partial_i \Longrightarrow \nabla_i \\
v^i \Longrightarrow u^i \\
h_i \Longrightarrow b_i \\
\tilde{\epsilon}_{ijk} \Longrightarrow \epsilon_{ijk}
\end{gather}
where $\epsilon_{ijk}$ is the Levi-Civita tensor (rather than symbol), defined by 
\begin{gather}
\epsilon_{ijk} = \sqrt{|g|} \tilde{\epsilon}_{ijk}.
\end{gather}

This implies that $\delta_{ij}^{\perp}$ will become $g_{ij} - b_i b_j$. It will be convenient to define a tensor $\beta_{ij}$ (following the notation from Krommes\cite{Krommes2018}) by 
\begin{gather}
\beta_{ij} \doteq - \epsilon_{ijk} b^k.
\end{gather}
Making these substitutions, and raising indices where necessary (since only an up-down pair of indices can be summed over), Eqs.~(\ref{eqn:W0}-\ref{eqn:W}) become
\begin{align}
^0W_{ab} &= \frac{3}{2} \big(b_a b_b - \frac{1}{3} g_{ab} \big) \big(b_m b_n - \frac{1}{3} g_{mn}\big) W^{mn} \label{eqn:W0tensorial}\\
^1W_{ab} &= \big(\delta^{\perp}_{am}  \delta^{\perp}_{nb} + \frac{1}{2} \delta^{\perp}_{ab} b_m b_n   \big) W^{mn}
\\
^2W_{ab} &= \big(\delta^{\perp}_{am}  b_b b_n + \delta^{\perp}_{nb} b_a b_m \big) W^{mn}\\
^3W_{ab} &= - \frac{1}{2} \big( \delta^{\perp}_{am}  \beta_{nb} - \delta^{\perp}_{nb} \beta_{am}   \big)W^{mn}\\
^4W_{ab} &= - \big(b_a b_m  \beta_{nb} + b_b b_n \beta_{am} \big) W^{mn}
\end{align}
and 
\begin{gather}
W_{ij} = \nabla_j u_i + \nabla_i u_j - \frac{2}{3} g_{ij} \nabla_k u^k . \label{eqn:Wtensorial}
\end{gather}
Eqs.~(\ref{eq:viscousSum}) and (\ref{eqn:W0tensorial}-\ref{eqn:Wtensorial}) completely define the viscous stress tensor, in a manifestly covariant way.

Now that we have the stress tensor, we can calculate the viscous force.
The generalization of the tensor divergence is again given by the covariant derivative.
Let $F$ be the conventional force vector, and $f$ be the force vector in covariant notation.
Then 
\beq
\mathbf{F} = \nabla \cdot \pi \Longrightarrow f^{j} = \nabla_i \pi^{ij},
\eeq 
where
\begin{align}
f^j = \nabla_i \pi^{ij} = \partial_i \pi^{ij} + \Gamma^i_{i\lambda} \pi^{\lambda j} +  \Gamma^j_{i\lambda} \pi^{i \lambda}. \label{eq:fT}
\end{align}
Our conventional vector components are finally given by converting this back from a contravariant vector (noting that $g_{ij}$ is diagonal by assumption):
\beq
F^i = \sqrt{g_{ii}} f^i .
\eeq
This gives a clear prescription for calculating the viscous force in an arbitrary coordinate system.
The sums can be carried out fairly straightforwardly in a symbolic math language such as Mathematica.
\begin{enumerate}
\item Start with the metric $g_{ij}$ that defines the coordinate system, and construct the inverse metric $g^{ij}$.
\item Construct the Christoffel symbols $\Gamma^{\nu}_{\mu \sigma}$ via Eq.~(\ref{eq:Christoffel}).
\item Convert velocity vectors $v^i$ and magnetic field vectors $h^i$ to covariant one-forms $u_i$ and $b_i$ via $u_i = v^i \sqrt{g_{ii}}$.
\item Construct the viscosity tensor given by Eqs.~(\ref{eq:viscousSum}) and (\ref{eqn:W0tensorial}-\ref{eqn:Wtensorial}). Indices are raised and lowered by applying the metric and inverse metric, e.g. $\te{W}{^m^n} = g^{ns} \te{W}{^m_s}$.
\item Calculate the contravariant force according to Eq.~(\ref{eq:fT}).
\item Convert the force back to orthonormal coordinates via $F^{i} = \sqrt{g_{ii}} f^i$.
\end{enumerate}

This method was used to calculate the viscous forces used in the text, using the metric for a cylindrical coordinate system:
\begin{equation}
    g_{ij} = \begin{pmatrix} 1&0&0\\0&r^2&0\\0&0&1\end{pmatrix}.
\end{equation}
In the next section, we explicitly calculate the forces in an alternate way, by converting the velocity vectors to Cartesian coordinates for substitution into the original Braginskii viscosity matrix.
The results agree. 

\subsection{Calculation in Cartesian Coordinates}

A straightforward way of calculating $(\nabla~\cdot~\pi)_r$ and $(\nabla~\cdot~\pi)_\theta$ directly from Braginskii's result is to first compute the Cartesian components of $\nabla \cdot \pi$ and then to convert the result to cylindrical coordinates. In this section we will use index notation but assume a Cartesian coordinate system.
Braginskii has 
\begin{gather}
W_{\alpha \beta} = \frac{\partial v_\alpha}{\partial x_\beta} + \frac{\partial v_\beta}{\partial x_\alpha} - \frac{2}{3} \delta_{\alpha \beta} \nabla \cdot \mathbf{v}
\end{gather}
and 
\begin{gather}
\pi_{zz} = - \eta_0 W_{zz} \\
\pi_{xx} = - \eta_0 \, \frac{W_{xx} + W_{yy}}{2} - \eta_1 \, \frac{W_{xx} - W_{yy}}{2} - \eta_3 W_{xy} \\
\pi_{yy} = -\eta_0 \, \frac{W_{xx} + W_{yy}}{2} - \eta_1 \, \frac{W_{yy} - W_{xx}}{2} + \eta_3 W_{xy} \\
\pi_{xy} = \pi_{yx} = - \eta_1 W_{xy} + \eta_3 \, \frac{W_{xx} - W_{yy}}{2} \\
\pi_{xz} = \pi_{zx} = - \eta_2 W_{xz} - \eta_4 W_{yz} \\
\pi_{yz} = \pi_{zy} = - \eta_2 W_{yz} + \eta_4 W_{xz}. 
\end{gather}
If $\mathbf{v} = v(r) \hat \theta$, then 
\begin{gather}
\partial_x v_x = - \frac{xy}{r^2} v' + \frac{xy}{r^3} v \\
\partial_x v_y = \frac{x^2}{r^2} v' + \frac{y^2}{r^3} v \\
\partial_y v_x = - \frac{y^2}{r^2} v' - \frac{x^2}{r^3} v \\
\partial_y v_y = \frac{xy}{r^2} v' - \frac{xy}{r^3} v
\end{gather}
and $\nabla \cdot \mathbf{v} = 0$, so 
\begin{gather}
W_{xx} = 2 \bigg( - \frac{xy}{r^2} v' + \frac{xy}{r^3} v \bigg) \\
W_{xy} = W_{yx} = \frac{x^2-y^2}{r^2} v' - \frac{x^2-y^2}{r^3} v \\
W_{yy} = 2 \bigg( \frac{xy}{r^2} v' - \frac{xy}{r^3} v \bigg) \\
W_{i z} = W_{z i} = 0. 
\end{gather}
Then we can evaluate the components of $\pi$. Let $\omega \doteq v/r$. Note that $\omega$ is sometimes used elsewhere to denote the vorticity $\nabla \times \bv$; these are equivalent in the case of solid-body rotation, but not for general rotation profiles. 
\begin{align}
\pi_{zz} &= 0 \\
\pi_{xx} &= - \pi_{yy} = \big[ \eta_1 (2xy) - \eta_3 (x^2-y^2) \big] \frac{\omega'}{r} \\
\pi_{xy} &= \pi_{yx} = \big[ - \eta_1 (x^2-y^2) - \eta_3 (2xy) \big] \frac{\omega'}{r} \\
\pi_{iz} &= \pi_{zi} = 0. 
\end{align}
The necessary derivatives of these components are 
\begin{align}
\frac{\partial \pi_{xx}}{\partial x} =& \, \bigg[ \eta_1' \, \frac{2 x^2 y}{r} - \eta_3' \, \frac{x (x^2-y^2)}{r} \bigg] \frac{\omega'}{r} \nonumber \\
&+ \big[ \eta_1 (2y) - \eta_3 (2x) \big] \frac{\omega'}{r} \nonumber \\
&+ \big[ \eta_1 (2xy) - \eta_3 (x^2-y^2) \big] \frac{x}{r} \bigg( \frac{\omega'}{r} \bigg)'
\end{align}
\begin{align}
\frac{\partial \pi_{yy}}{\partial y} =& \, \bigg[ - \eta_1' \, \frac{2xy^2}{r} + \eta_3' \, \frac{y (x^2-y^2)}{r} \bigg] \frac{\omega'}{r} \nonumber \\
&+ \big[ - \eta_1 (2x) - \eta_3 (2y) \big] \frac{\omega'}{r} \nonumber \\
&+ \big[ - \eta_1 (2xy) + \eta_3 (x^2 - y^2) \big] \frac{y}{r} \bigg( \frac{\omega'}{r} \bigg)'
\end{align}
\begin{align}
\frac{\partial \pi_{xy}}{\partial y} =& \, \bigg[ - \eta_1' \, \frac{y (x^2 - y^2)}{r} - \eta_3' \, \frac{2xy^2}{r} \bigg] \frac{\omega'}{r} \nonumber \\
&+ \big[ \eta_1 (2 y) - \eta_3 (2 x) \big] \frac{\omega'}{r} \nonumber \\
&+ \big[ - \eta_1 (x^2-y^2) - \eta_3 (2xy) \big] \frac{y}{r} \bigg( \frac{\omega'}{r} \bigg)'
\end{align}
\begin{align}
\frac{\partial \pi_{yx}}{\partial x} =& \, \bigg[ - \eta_1' \, \frac{x (x^2 - y^2)}{r} - \eta_3' \, \frac{2 x^2 y}{r} \bigg] \frac{\omega'}{r} \nonumber \\
&+ \big[ - \eta_1 (2x) - \eta_3 (2y) \big] \frac{\omega'}{r} \nonumber \\
&+ \big[ -\eta_1 (x^2-y^2) - \eta_3 (2xy) \big] \frac{x}{r} \bigg( \frac{\omega'}{r} \bigg)' . 
\end{align}
The components of $\nabla \cdot \pi$ can be written as: 
\begin{align}
(\nabla \cdot \pi)_x &= \pi_{xx,x} + \pi_{xy,y} \\
(\nabla \cdot \pi)_y &= \pi_{yy,y} + \pi_{yx,x} 
\end{align}
and 
\begin{align}
(\nabla \cdot \pi)_r &= \frac{x (\nabla \cdot \pi)_x + y (\nabla \cdot \pi)_y}{r} \\
(\nabla \cdot \pi)_\theta &= \frac{-y (\nabla \cdot \pi)_x + x (\nabla \cdot \pi)_y}{r}. 
\end{align}
All that remains is to evaluate these. 
\begin{align}
(\nabla \cdot \pi)_x &= y \eta_1' \omega' - x \eta_3' \omega' \nonumber \\
&\hspace{5 pt}+ \frac{4y}{r} \, \eta_1' \omega' - \frac{4x}{r} \, \eta_3' \omega' \nonumber \\
&\hspace{5 pt}+ y r \eta_1 \bigg( \frac{\omega'}{r} \bigg)' - x r \eta_3 \bigg( \frac{\omega'}{r} \bigg)'
\end{align}
\begin{align}
(\nabla \cdot \pi)_y &= - x \eta_1' \omega' - y \eta_3' \omega' \nonumber \\
&\hspace{5 pt}- \frac{4x}{r} \, \eta_1 \omega' - \frac{4y}{r} \, \eta_3 \omega' \nonumber \\
&\hspace{5 pt} - x r \eta_1 \bigg( \frac{\omega'}{r} \bigg)' - y r \eta_3 \bigg( \frac{\omega'}{r} \bigg)'. 
\end{align}
Then in cylindrical coordinates, 
\begin{align}
(\nabla \cdot \pi)_r &= - r \eta_3' \omega' - 4 \eta_3 \omega' - r^2 \eta_3 (\omega'/r)' \\
&= - r \eta_3' \omega' - 3 \eta_3 \omega' - r \eta_3 \omega'' \\
&= - \frac{1}{r^2} \big( r^3 \omega' \eta_3 \big)'
\end{align}
\begin{align}
(\nabla \cdot \pi)_\theta &= - \frac{1}{r^2} \big( r^3 \omega' \eta_1 \big)' .
\end{align}
The azimuthal expression is the same as the radial expression, just exchanging $\eta_3$ with $\eta_1$. It can be rewritten as 
\begin{gather}
(\nabla \cdot \pi)_\theta = - \eta_1 \frac{\partial}{\partial r} \bigg[ \frac{1}{r} \frac{\partial (r v)}{\partial r} \bigg] - \frac{\partial \eta_1}{\partial r} r \frac{\partial}{\partial r} \bigg( \frac{v}{r} \bigg). 
\end{gather}

\end{appendix}

\bibliographystyle{apsrev4-1} 
\bibliography{spinBib.njf.bib}

\end{document}